\documentclass[aps,prl, twocolumn,linenumber,floatfix,citeautoscript,superscriptaddress]{revtex4-1}
\usepackage[utf8]{inputenc}
\usepackage{geometry}
\usepackage{setspace}
\usepackage{times}
\usepackage[T1]{fontenc}
\usepackage{amsmath}
\usepackage{hyperref} 
\hypersetup{colorlinks=true,linkcolor=blue,anchorcolor=blue,citecolor=blue,urlcolor=black}
\usepackage{indentfirst}
\usepackage{graphicx}
\usepackage{booktabs}
\usepackage{makecell}
\usepackage{gensymb}
\usepackage{verbatim}
\usepackage{booktabs}
\usepackage{threeparttable}
\usepackage{multirow}
\usepackage{lineno}
\usepackage{footnote}   
\usepackage{graphicx}   
\usepackage{subcaption} 
\usepackage{appendix}   
\usepackage{amssymb}
\usepackage{mathptmx}
\usepackage{bm}

\newcommand{\IHEP}{\affiliation{Institute~of~High~Energy~Physics, Chinese Academy of Sciences, Beijing, China}}
\newcommand{\UCAS}{\affiliation{University of Chinese Academy of Sciences, Beijing, China}}
\newcommand{\SDU}{\affiliation{Shandong University, Jinan, China}}
\newcommand{\IMP}{\affiliation{Institute of Modern Physics, Chinese Academy of Sciences, Lanzhou, China}}

\begin{document}

\title{Impact of the $^5$Li resonance in $\alpha$-$p$ elastic scattering on precision measurements of neutrino oscillation parameters}

\author{Mingyuan Wang}\UCAS\IHEP
\author{Yiyang Li}\IMP\UCAS
\author{Suqing Hou}\IMP
\author{Fei Xiao}\UCAS\IHEP
\author{Yaoguang Wang}\altaffiliation{Corresponding author:wangyaoguang@sdu.edu.cn}\SDU
\author{Zeyuan Yu}\altaffiliation{Corresponding author: yuzy@ihep.ac.cn}\IHEP

\date{\today}

\begin{abstract}

Precision measurements of four neutrino oscillation parameters—$\theta_{12}$, $\theta_{13}$, $\Delta m^2_{21}$, and |$\Delta m^2_{31}$|—face significant interference from a previously overlooked correlated background. Recent findings from the SNO+ and JUNO experiments reveal that cascade decays of $^{214}$Bi-$^{214}$Po in liquid scintillator detectors can mimic inverse beta decay signals from reactor and geoneutrinos, with a misidentification probability on the order of $10^{-4}$ when hydrogen neutron capture is used—a rate ten times higher than Geant4 simulations predicted.

This work identifies the $^5$Li resonance in $\alpha$-$p$ elastic scattering as the underlying cause. For alpha energies above 5~MeV, the cross section is hundreds of times larger than that of Rutherford scattering. After correctly incorporating the differential cross section into Geant4, the misidentification probability is recalculated as 1.9$\times$10$^{-4}$. The simulated shape of the long tail in the alpha deposited energy also differs from the extrapolation models currently used by SNO+ and JUNO.

These results will assist both experiments in more accurately estimating this novel background, thereby refining measurements of neutrino oscillation parameters and the geoneutrino flux. Additionally, the study implies an overlooked background with a rate of 0.5 events per detector per day in the Daya Bay $\theta_{13}$ analysis using hydrogen neutron capture, leading to an increase of $\sin^22\theta_{13}$ by approximately 0.012. Consequently, the Particle Data Group’s reported $\sin^2\theta_{13}$ value shall increase by about 0.006~(1$\sigma$).

\end{abstract}
\maketitle

$Introduction$ Neutrino oscillation studies have now entered an era of precision measurement. The determination of the four key oscillation parameters—$\theta_{12}$, $\theta_{13}$, $\Delta m^2_{21}$, and |$\Delta m^2_{31}$|—is currently or will soon be dominated by reactor neutrino experiments employing liquid scintillator (LS) detectors.
These parameters are fundamental for testing the three-neutrino mixing paradigm and for the determination of the neutrino mass ordering~\cite{An:2015jdp,JUNO:2025gmd} and searching for CP violation in neutrino oscillations~\cite{T2K:2025wet,Hyper-Kamiokande:2025fci,DUNE:2020jqi}.

Reactor antineutrinos are primarily detected via the inverse beta decay (IBD) process, which produces a distinctive correlated signal: a prompt positron annihilation followed by a delayed neutron capture on hydrogen~(nH) or gadolinium~(nGd). This coincidence signature suppresses backgrounds from natural radioactivity by several orders of magnitude.
Consequently, the accurate identification and estimation of intrinsic correlated backgrounds—such as those from cosmogenic isotopes ($^9$Li and $^8$He)~\cite{DayaBay:2024xye,RENO:2022xbr,DoubleChooz:2018kvj,KamLAND-Zen:2023spw} and radiogenic processes—remain crucial.

Two types of radiogenic processes, $\alpha$-$n$ and $\alpha$-$p$ are documented in the literature.
In LS, the neutron yield per alpha particle from natural radioactive decays is typically between $10^{-7}$ and $10^{-8}$~\cite{Zhao:2013mba}, making it negligible when LS radiopurity is well controlled.
A novel correlated background induced by $\alpha$-$p$ elastic scattering process has recently been identified by advanced LS detectors such as JUNO\cite{JUNO:2025gmd} and SNO+~\cite{SNO:2025koj}.
Both experiments report a probability on the order of $10^{-4}$ for a $^{214}$Bi-$^{214}$Po cascade decay to mimic nH IBD.
The prompt signal originates from the beta decay of $^{214}$Bi with an end point energy of 3.2~MeV.
The delayed signal corresponds to the long tail of the visible energy of 7.68~MeV $\alpha$ particle from $^{214}$Po decay.
The tail is explained as the energy deposit from scattering protons which feature a 25\% quenching effect in LS than alphas.
Due to the comparable lifetimes of $^{214}$Po decay (238~$\mu$s) and neutron capture on hydrogen (220~$\mu$s), a $^{214}$Po alpha particle with visible energy near that of a neutron capture event will cause the $^{214}$Bi–$^{214}$Po cascade decay to easily mimic IBDs.

The probability predicted by Geant4 simulations using the default packages~\cite{Allison:2016lfl} is on the order of $10^{-5}$, which is ten times smaller than the experimental values. This background is critically important because the prompt signal from $^{214}$Bi, which has an energy spectrum peaked at 2.5~MeV, overlaps with the prompt energy peaks of both reactor neutrinos and geo-neutrinos originating from terrestrial $^{214}$Bi decays. Therefore, it is essential to understand the long tail of the alpha energy deposition in LS detectors and to address the discrepancy between data and simulation. Furthermore, a careful reassessment of this background in past $\theta_{13}$ and geo-neutrino experiments is warranted.


$Review~of~experimental~and~simulation~results$
The rates of $^{214}$Bi–$^{214}$Po decays and the corresponding IBD background rates in JUNO and SNO+ are summarized in Table~\ref{table1}.  The high-energy tail of the $\alpha$ particle energy deposit is modeled with either a linear or an exponential function and fitted to the measured $^{214}$Po or $^{215}$Po spectrum~\cite{JUNO:2025gmd,SNO:2025koj}. The background rate in the IBD sample is then obtained by extrapolating the fitted function into the delayed energy region of IBD events (2–2.5~MeV).
JUNO and SNO+ report that each $^{214}$Bi–$^{214}$Po decay can mimic an IBD event with a probability of 0.014\% and 0.024\%, respectively. These probabilities are about 2,000 times higher than those expected from $\alpha$–$n$ reactions.

\begin{table}[!htb]
\caption{\label{table1} Published data on $^{214}$Bi–$^{214}$Po decay rates and  corresponding IBD background from JUNO and SNO+. For comparison, we also provide two sets of Geant4 simulation results, with the latter~(Geant4-m) including $^5$Li resonance. For $^{212}$Po decays in the $^{232}$Th chain, the simulation results are provided. }
    \centering
    \begin{tabular}{c|c|c|c}\hline
          & $^{214}$Po  & IBD bkg & Probability \\ \hline
        JUNO &  1300/day & 0.18$\pm$0.1/day & (0.014$\pm$0.008)\% \\ \hline
        SNO+  & 28776 & 7$\pm$6 & (0.024$\pm$0.02)\%\\ \hline
        Geant4 & 10$^7$ & 100 & 0.001\% \\ \hline
        Geant4-m & 10$^7$ & 1900 & 0.019\% \\ \hline
        & $^{212}$Po  & IBD bkg & Probability \\ \hline
        Geant4 & 10$^7$ & 300 & 0.003\% \\ \hline
        Geant4-m & 10$^7$ & 5900 & 0.059\% \\ \hline
    \end{tabular}
\end{table}

To understand the long tail in the alpha energy spectrum in LS, simulation studies are performed based on Geant4.
The LS was configured with a density of 0.86 g/cm$^3$ and a hydrogen mass fraction of 0.12.
Several factors influence the results of $\alpha$-particle simulation, such as the limit of step length, the production cuts of secondary particles, the quenching factor, and the $\alpha$-$p$ cross section.

The quenching of $\alpha$-particles and protons is a crucial effect in LS-based detectors. Typically, this quenching is described by Birks’ law. While numerous measurements of the quenching effect have been reported in the literature, often expressed in terms of Birk’s constants ($k$B, $k$C), these constants are generally obtained by comparing simulated quenching curves with experimental data.
Consequently, their values depend on simulation configurations, such as step length and production cuts. To date, there is no unified quenching model or single set of parameters that can accurately describe all experimental data.
For linear alkylbenzene (LAB)-based LS, we adopt a proton quenching parameter of $k$B=7.5$\times10^{-3}$ g/cm$^2$/MeV from Ref.~\cite{Yang:2018wje} for protons.
For $\alpha$-particles we chose $k$B=3.5$\times10^{-3}$ g/cm$^2$/MeV by tuning the visible energy of $^{214}$Po peak to approximate 0.9~MeV, consistent with Daya Bay data in Fig.~30 in Ref.~\cite{DayaBay:2016ggj}.

In Geant4, alpha particles undergo two distinct elastic scattering processes: Rutherford scattering, handled by the MSC model or optionally the G4ScreenedNuclearRecoil model if proton recoil is of interest, and hadronic elastic scattering. The cross section for the former is well established, whereas that for the latter is computed using the Glauber–Gribov model and about 400~mb for a 7~MeV alpha. Furthermore, final-state kinematics are generated by the GHEISHA model, which tends to produce proton scattering preferentially around $90^{\circ}$.

The step-length limit and production cuts for secondary particles in Geant4 influence both the quenching model and the number of generated secondaries, as also noted in previous studies (e.g., Refs.~\cite{Aberle:2011zz,DayaBay:2019fje}). In our simulation, we set the nominal production cut for $e^-/e^+$ to 0.1~mm and varied the proton production cut across several values. A dedicated test was also performed by imposing an upper limit on the energy loss per step. No significant variation was observed in the results across these configurations, primarily because the applicable energy range for Rutherford scattering in Geant4 extends down to 50~keV, which is sufficiently low for the processes considered here.

\begin{figure}[!htb]
\includegraphics[width=1\linewidth]{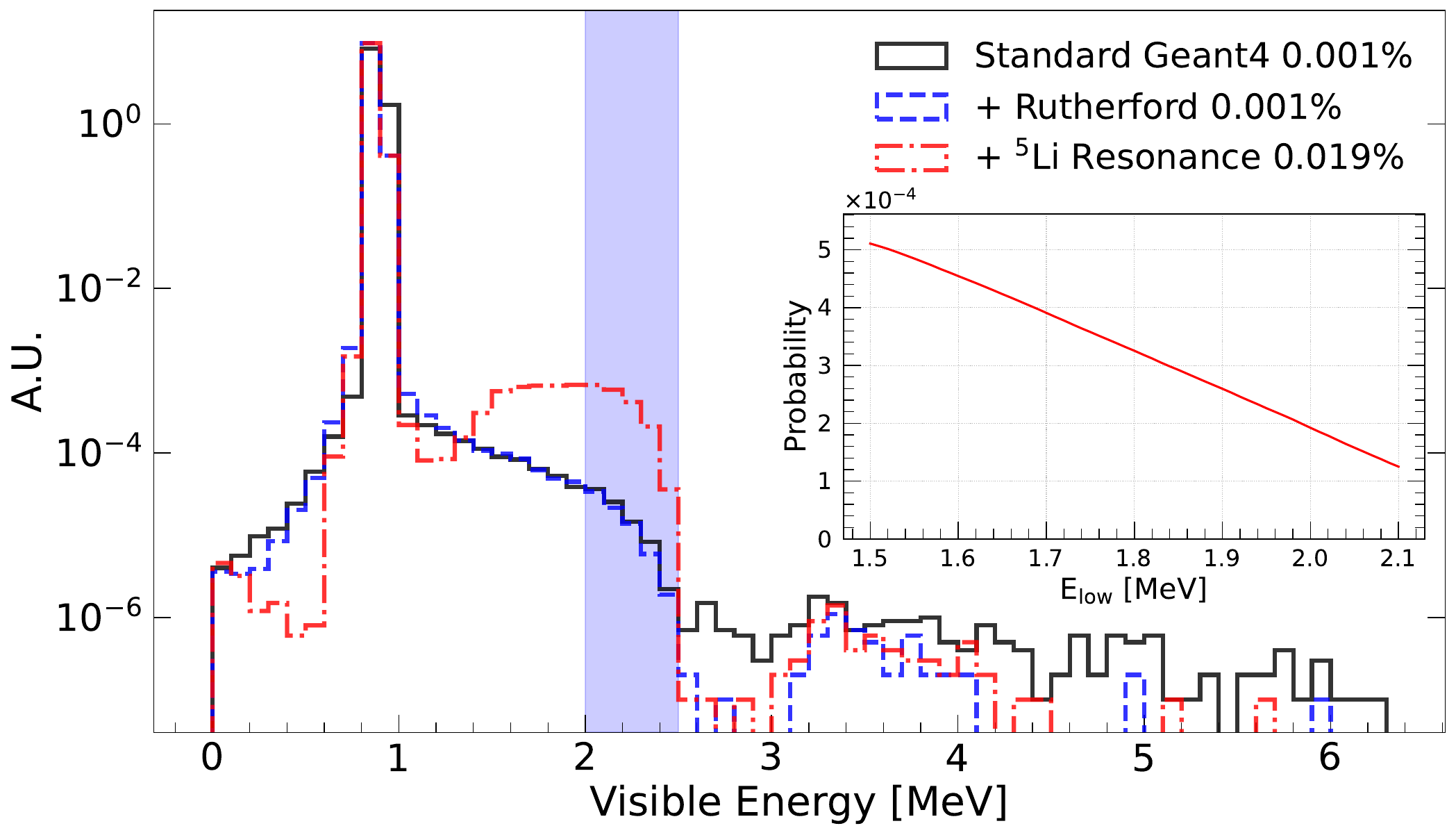}
\caption{Simulated distributions of visible energy from $^{214}$Po alpha with default Geant4 models, Rutherford scattering alone, and $^5$Li resonance alone. The inlet plot shows the integral probability between $E_{low}$ to 2.5~MeV for the $^5$Li resonance line. }
\label{fig1}
\end{figure}

\begin{figure}[!htb]
    \centering
    \includegraphics[width=1\linewidth]{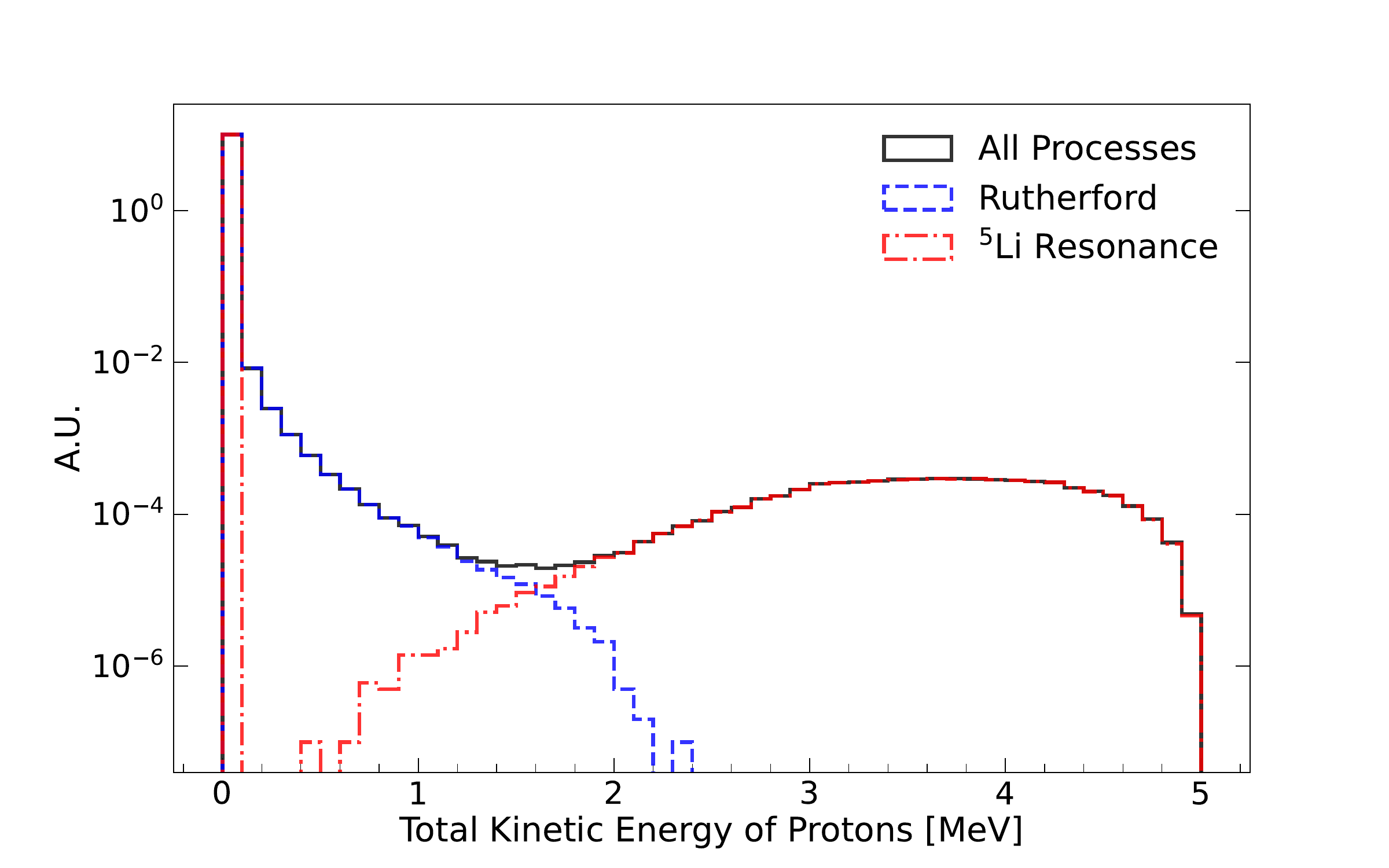}
    \caption{Summed kinetic energy of scattered protons per $^{214}$Po alpha. Contributions from Rutherford scattering and $^5$Li resonance are shown.}
    \label{fig2}
\end{figure}

Using the simulation setup described above, the visible energy spectrum of $^{214}$Po alpha particles~(7.68~MeV) is shown in Fig.~\ref{fig1}.
Only about $1.5 \times 10^{-5}$ of these alpha particles deposit a visible energy in the 2–2.5~MeV window, which could mimic the 2.2~MeV gamma from neutron capture on hydrogen.
Rutherford scattering alone produces protons with kinetic energies below 2~MeV as shown in Fig.~\ref{fig2}. Given that the visible energy for protons in this range corresponds to only about 30\% of their kinetic energy, it cannot reproduce the long tail extending up to 3~MeV.

$^5Li~resonance$ In the MeV energy range, $\alpha$–$p$ scattering is governed by distinct mechanisms: Rutherford scattering dominates for $\alpha$-particle energies below 4~MeV, while above this threshold the $^5$Li resonance becomes the predominant process~\cite{PUSA2004686}.
In nuclear physics, $\alpha + p \rightarrow {}^5\mathrm{Li} \rightarrow \alpha + p$ represents a classic resonant scattering process. The resonance occurs at an incident $\alpha$-particle energy of 8.8~MeV and has a width of approximately 1.5~MeV.
Using the $R$-matrix formalism, we fitted experimental differential cross-section data~\cite{PUSA2004686} and derived the corresponding angular distributions of the scattered protons.

\begin{figure}[!htb]
\includegraphics[width=1\linewidth]{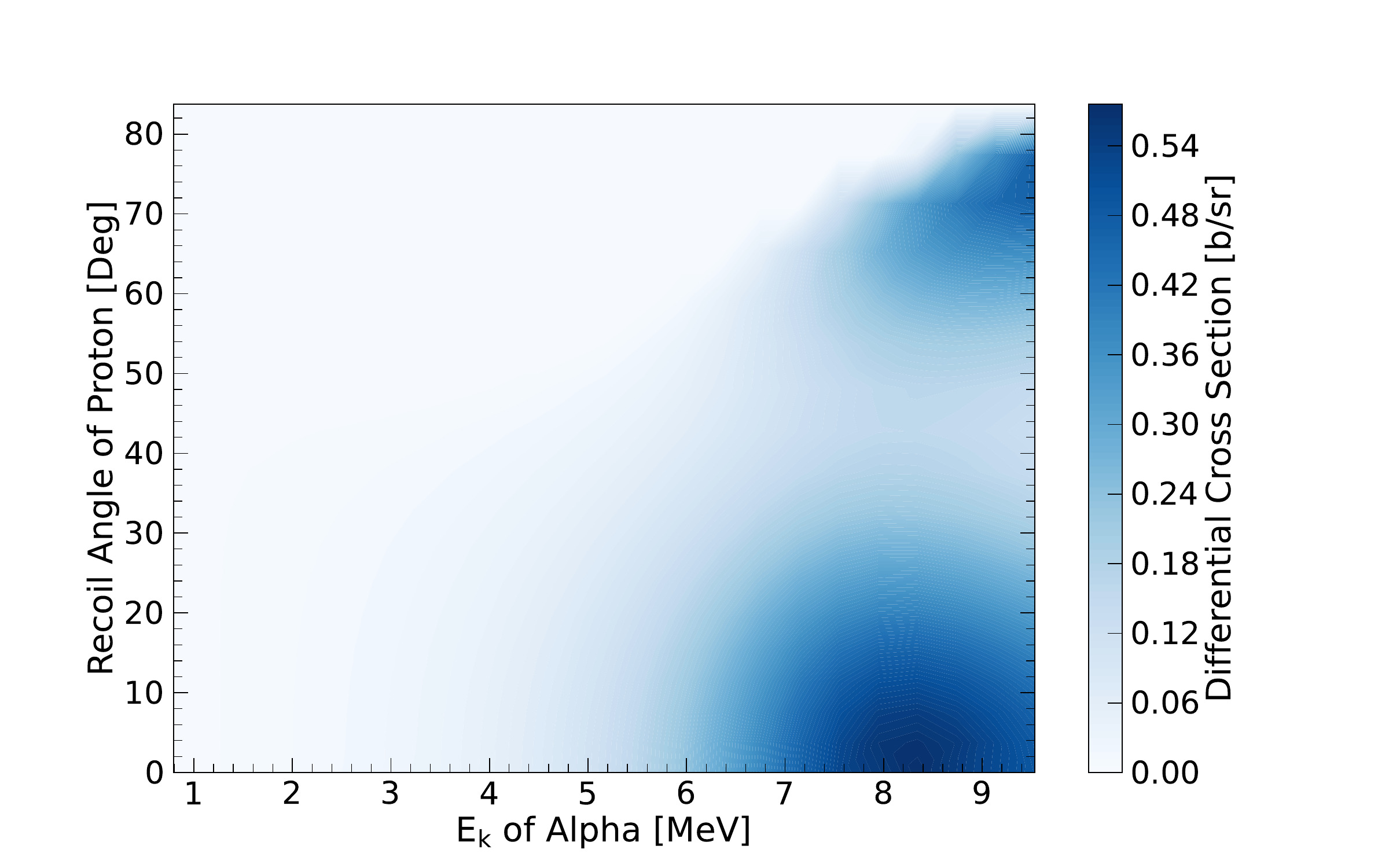}
\caption{Differential cross section of $\alpha$-$p$ elastic scattering for the proton scattering angle from 0$^{\circ}$ to $90^{\circ}$. The top left region is dominated by Rutherford scattering and the bottom right region by $^5$Li resonance.}
\label{fig3}
\end{figure}

Fig.~\ref{fig3} compares the angular-dependent differential cross sections for Rutherford scattering and for the $^5$Li resonance. The resonant process strongly favors small scattering angles, transferring a larger fraction of the total kinetic energy to the proton compared to Rutherford scattering. This leads to a pronounced population of high-energy scattered protons, as shown in Fig.~\ref{fig2}.
After quenching, these energetic protons produce the long tail observed in the visible energy spectrum of alpha particles in Fig.~\ref{fig1}. The probability for a $^{214}$Po alpha to deposit visible energy in the 2–2.5 MeV window is about 0.019\%, which aligns with the order of magnitude reported by SNO+ and JUNO. Moreover, the shape of this long tail differs from a linear or exponential function due to the kinematics in the reaction. It suggests that JUNO and SNO+ can better estimate this background using the simulated shape.
The integrated probabilities starting from different energies to 2.6~MeV are also provided in the inlet of
Fig.~\ref{fig1}.

Beyond $^{214}$Po, the $^{212}$Bi–$^{212}$Po cascade can also generate an IBD background through $\alpha$–$p$ scattering.
The alpha particle from $^{212}$Po (8.78~MeV) lies closer to the $^5$Li resonance region than the 7.69~MeV alpha from $^{214}$Po decay.
Simulations in Geant4 using the original and updated cross sections yield probabilities of 0.003\% and 0.059\%, respectively, for a $^{212}$Po alpha with visible energy between 2~MeV and 2.5~MeV.
Nevertheless, the 439~ns lifetime of $^{212}$Po is typically shorter than the electronics readout window (0.5 to 1~$\mu$s), causing about 70\% to 90\% of $^{212}$Bi–$^{212}$Po cascade decays to be recorded within the same event, not two separated events, and reducing the probability to be identified as an IBD event.

$Impact~on~\theta_{13}$ Table~\ref{table2} summarizes the relevant published information from reactor neutrino and geoneutrino experiments.
\begin{itemize}
    \item Daya Bay reported $^{214}$Bi–$^{214}$Po and $^{212}$Bi–$^{212}$Po rates in GdLS, but the corresponding values in pure LS were not provided (marked as N/F in the table)~\cite{DayaBay:2016ggj}.
    \item RENO did not report specific natural radioactivity levels, but stated that the $\alpha$-$n$ background is negligible in their analysis~\cite{RENO:2025wbu}.
    \item Double Chooz reported the daily rates of $^{214}$Bi–$^{214}$Po and $^{212}$Bi–$^{212}$Po in the full detector volume ~\cite{Hofmann:2012uug}. Spatial distributions were also reported. $^{212}$Bi–$^{212}$Po concentrated in GdLS region while $^{214}$Bi–$^{214}$Po in LS is about 2 times higher than GdLS.
\end{itemize}

According to the delayed energy cut of 1.9~MeV to 2.6~MeV in Daya Bay nH analysis~\cite{2016nH}, under the assumption of 100\% detection efficiency, the probability of a $^{214}$Po and $^{212}$Po to form an IBD background is determined to 0.026\% and 0.079\%. Thus, in the GdLS target, the expected rates for the $^{214}$Bi–$^{214}$Po and $^{212}$Bi–$^{212}$Po induced IBD background are 0.2/day and 13.6/day, respectively.
To compare these rates with reactor neutrino signals, the efficiencies of two key selection cuts must be considered: the prompt energy cut (1.5–12~MeV) and the time interval cut (>1$\mu$s).
For $^{214}$Bi–$^{214}$Po, the efficiency exceeds 90\%, due to the 3.2~MeV endpoint energy of $^{214}$Bi beta decay and the 238~$\mu$s lifetime of $^{214}$Po.
For $^{212}$Bi–$^{212}$Po, the 1.5~MeV prompt energy cut removes about 90\% of $^{212}$Bi events, while the 1~$\mu$s time interval cut removes approximately 90\% of $^{212}$Po events, leading to a rate of 0.13/day.
In the 22~ton LS region, the same $^{214}$Bi–$^{214}$Po rate as in the GdLS region is assumed, yielding a background rate of 0.22/day before efficiency corrections.
Overall, the Daya Bay nH analysis finds a total background rate of 0.5/day/detector, which corresponds to a background to signal ratio of 0.1\% and 0.86\% in the near and far detectors, respectively.

The influence of this backgrounds on the measurement of $\theta_{13}$ was estimated using a simple conversion between the near–far rate deficit and $\sin^2 2\theta_{13}$.
A near-far rate deficit ratio 0.950 corresponded to a $\sin^2 2\theta_{13}$ value of 0.071~\cite{DayaBay:2016ziq}.
A deficit of 0.940 corresponded to $\sin^2 2\theta_{13}$ of 0.092~\cite{DayaBay:2012fng}.
It indicates that each 0.01 rate deficit is proportional to an approximately 0.015 change on the central value of $\sin^2 2\theta_{13}$.
The 0.86\% background over signal ratio in far detectors will reduce the deficit by 0.008, resulting in a 0.012 increase for $\sin^2 2\theta_{13}$.

The latest $\sin^2 2\theta_{13}$ value from Daya Bay nH analysis is 0.076$\pm$0.005~\cite{DayaBay:2024hrv}. It will increase to about 0.088 if the Bi-Po background is properly considered.
The $\sin^2 \theta_{13}$ with a central value of 0.0216$\pm 0.006$ reported in PDG 2025~\cite{ParticleDataGroup:2024cfk} is the error weighted average of several measurements, including the Daya Bay nH analysis. After considering the Bi-Po background, the central value 0.0216 will increase by 1$\sigma$ to 0.0222.

For Double Chooz, the prompt energy cut was 0.7~MeV and the minimal time interval cut is 0.5~$\mu$s.
About 15\% $^{212}$Bi–$^{212}$Po background can pass these cuts.
It means a total of 0.1/day background, consisting of 0.09/day from $^{214}$Bi–$^{214}$Po and 0.01/day from $^{212}$Bi–$^{212}$Po.
The background over signal ratio is smaller than 0.1\% in the far detector, which puts little impact on their oscillation measurements.

\begin{table*}[!hbt]
    \centering
    \begin{tabular}{c|c|c|c|c}
    \hline
          & $^{214}$Po GdLS & $^{212}$Po GdLS & $^{238}$U LS& $^{232}$Th LS\\ \hline
        Daya Bay~\cite{DayaBay:2016ggj} & 0.009Hz & 0.2Hz &N/F &N/F \\ \hline
        Double Chooz~\cite{Hofmann:2012uug} & 150/day & 300/day & 300/day & 0 \\ \hline
        RENO &N/F & N/F& N/F& N/F\\ \hline
        KamLAND~\cite{KamLAND:2002uet} & N/A & N/A & 3.5$\times10^{-18}$g/g & 5.2$\times10^{-17}$g/g \\ \hline
        KamLAND-$^7$Be~\cite{KamLAND:2014gul} & N/A & N/A & 5.0$\times10^{-18}$g/g & 1.3$\times10^{-17}$g/g \\ \hline
        Borexino-solar~\cite{Borexino:2013zhu} & N/A & N/A & 5.3$\times10^{-18}$g/g & 3.8$\times10^{-18}$g/g \\ \hline
        Borexino-geo~\cite{Borexino:2019gps} & N/A& N/A &1.1$\times10^5$ decays & N/F \\ \hline
        JUNO~\cite{JUNO:2025fpc} & N/A & N/A & 7.5$\times10^{-17}$g/g & 8.2$\times10^{-17}$g/g \\ \hline
        SNO+~\cite{SNO:2024wzq} & N/A & N/A & 5$\times10^{-17}g/g$& 5$\times10^{-17}$g/g\\ \hline
    \end{tabular}
    \caption{Published information from reactor, geo and solar neutrino experiments. $^{238}$U and $^{232}$Th levels are measured with $^{214}$Bi–$^{214}$Po and $^{212}$Bi–$^{212}$Po, respectively, under the assumption of secular equilibrium. N/F means not found in literature. }
    \label{table2}
\end{table*}

$Impact~on~geo$-$neutrinos$ KamLAND, Borexino, JUNO, and SNO+ are LS based experiments. They provided contamination levels for both the $^{238}$U and $^{232}$Th chains, determined from the $^{214}$Bi–$^{214}$Po and $^{212}$Bi–$^{212}$Po cascade decays measured during stable periods, under the assumption of secular equilibrium.

In the KamLAND experiment~\cite{Gando:2013nba}, the background from the $^{214}$Bi-$^{214}$Po cascade in the $^{238}$U decay chain during stable periods contributes only about 1\% to the observed geoneutrino signal of $116^{+28}_{-27}$ events. However, as noted in Ref.~\cite{KamLAND:2009ply}, $^{222}$Rn could be introduced into the detector during calibration procedures. Although the exact amount of $^{222}$Rn was not clearly reported, the rate of $^{214}$Bi–$^{214}$Po events was observed to increase by a factor of about 100 following an off-axis calibration. Fortunately, $^{222}$Rn decays with a half-life of 5.5 days. It would therefore be valuable for the KamLAND experiment to re-examine the total number of $^{214}$Bi–$^{214}$Po decays in their dataset.

In the case of Borexino’s geoneutrino analysis~\cite{Borexino:2019gps}, data taken during the water extraction period, which exhibited higher $^{222}$Rn rates, were included. The total number of $^{214}$Bi–$^{214}$Po decays observed in that dataset was reported.
The Borexino's pseudocumene based LS and lower concentration of fluorescence~(1.5~g/L PPO) lead to about 10\% heavier quenching for alpha particles than Daya Bay.
We don't have enough information to simulate the quenching effect and the pulse shape efficiency in Borexino's study.
This can be further done by the collaboration.

The beta spectrum of $^{214}$Bi partially overlaps with the geoneutrino spectrum from the $^{238}$U decay chain.
It means that the $^{214}$Bi–$^{214}$Po background not only influences measurements of the total geoneutrino flux, but also those of the U/Th ratio in the Earth. The geoneutrino experts have been aware of this background.

$Summary$ A classical nuclear process has been found to put a significant influence on neutrino oscillation experiments.
By incorporating the correct differential cross section for $\alpha$–$p$ scattering into Geant4, the newly identified $^{214}$Bi–$^{214}$Po background in JUNO and SNO+ is well explained by the $^5$Li resonance process.
The simulation also provides a more accurate energy shape model, enabling these experiments to improve their data-driven estimation of this background. This will lead to more precise future measurements of neutrino oscillations and geo-neutrinos.
Regarding the mixing angle $\theta_{13}$, while measurements based on neutron capture on gadolinium remain unaffected, the value of $\sin^2 2\theta_{13}$ derived from the Daya Bay nH analysis is underestimated by approximately 0.012.
More detailed studies by the collaborations are expected in the near future.

\section{Acknowledgement}
The work is supported by the National Key R\&D Program of China under Grant (2023YFA1606102 and N2022YFA1603300), the Research Program of State Key Laboratory of Heavy Ion Science and Technology, In-
stitute of Modern Physics, Chinese Academy of Sci-
ences, under Grant No.HIST2025CS05, the National
Natural Science Foundation of China (Grant
No.12547111).

\bibliographystyle{apsrev}
\bibliography{AlphaP}
\end{document}